\documentclass{article}
\usepackage{nips13submit_e,times}
\usepackage[german,american]{babel}
\usepackage[a4paper,top=3.5cm,bottom=3.5cm,left=3.5cm,right=3.5cm,marginparwidth=1.75cm]{geometry}
\usepackage{amsmath}
\usepackage{amsfonts}
\usepackage{amssymb}
\usepackage{graphicx}
\usepackage[font=small,labelfont=bf]{caption}
\usepackage{subcaption}
\usepackage{float}
\usepackage[OMLmathsfit]{isomath}
\usepackage{amssymb} 
\usepackage{amsxtra} 
\usepackage{bm} 
\usepackage{mathtools}
\usepackage{verbatim}
\usepackage[colorlinks=true, allcolors=blue]{hyperref}
\nipsfinalcopy 
\DeclareCaptionLabelFormat{mylabel}{#1 #2\hspace{1.5ex}}
\usepackage[numbers,sort&compress]{natbib}

\definecolor{darkgreen}{rgb}{0.0,0.5,0.0}
\definecolor{BurntOrange}{rgb}{0.8,0.3,0.0}
\definecolor{mygray}{gray}{0.5}

\def\msbi#1{\mathsfbfit{#1}}

\newcommand{\figref}[1]{Fig.~\ref{#1}}
\newcommand{\equref}[1]{Eq.~(\ref{#1})}
\newcommand{\secref}[1]{Sec.~(\ref{#1})}

\providecommand{\keywords}[1]
{
  {
  \small	
  \textbf{\textit{Keywords---}} #1
}}

\title{\textbf{Teaching Solid Mechanics to Artificial Intelligence: \\ a fast solver for heterogeneous solids}}

\author{Jaber Rezaei Mianroodi$^{1,*}$, Nima H. Siboni$^{2}$, Dierk Raabe$^{1}$ \\
       {\footnotesize $^1$Microstructure Physics and Alloy Design,} \\ 
        {\footnotesize Max-Planck-Institut f\"ur Eisenforschung, D\"usseldorf, Germany} \\
        {\footnotesize $^2$DeepMetis, Lohm\"uhlenstraße 65, 12435 Berlin, Germany} \\
       {\footnotesize $^*$Corresponding Author j.mianroodi@mpie.de}
} 

\begin{document} 
\maketitle

\begin{abstract} 
We propose a deep neural network (DNN) as a fast surrogate model for local stress (and in principle strain) calculation in inhomogeneous non-linear material systems. We show that the DNN predicts the local stresses with about 3.8\% mean absolute percentage error (MAPE) for the case of heterogeneous elastic media and a mechanical phase contrast of up to factor of 1.5 among neighboring domains, while performing 103 times faster than spectral solvers. The speed-up arises from the fact that after training, the DNN predicts the stress without any iterations, as opposed to the iterative nature of standard non-linear solvers. 
The new DNN surrogate model also proves suited for general purposes: it is capable to reproduce the stress distribution in geometries topologically far different from those used for training, implying effective learning of scenarios described by the underlying partial differential equations. 
Even in the case of elasto-plastic materials with up to 2 times mechanical contrast in elastic stiffness and 4 times in yield stress among adjacent regions, where conventional solvers typically require a substantial number of iterations to arrive at stress predictions, the trained model simulates  the micromechanics with a MAPE of 6.4\% in  one single  forward evaluation step of the network, i.e. without any  iterative calculations even for the case of such a non-linear problem. The results reveal a completely new and highly efficient approach to solve non-linear mechanical boundary value problems and / or augment existing solution methods in corresponding hybrid variants, with an acceleration up to factor of 8300 for heterogeneous elastic-plastic materials in comparison to the currently fastest available solvers. 
\end{abstract} 

\section{Introduction}

The mechanical response of materials depends highly  on the microstructure and its heterogeneity, including all defects, phases and chemical features. An inseparable task in  modeling a corresponding coupled multi-physics material problem that links mechanical properties to microstructure, consists in  solving the underlying differential equations for mechanical equilibrium. This task is particularly challenging for the case of non-linear material response,  highly inhomogeneous material properties and complex microstructure topologies. Several methods to numerically solve such non-linear solid mechanics problems have already been developed and are in use, such as spectral solvers and the finite element method. However, ever-increasing sophistication of material models by including more physics and coupled problems in multi-scale methods is bringing these numerical solvers to their limits. 

Therefore, a number of attempts were made to speed up material modeling using artificial intelligence (AI) in general, and specifically using deep neural networks. For example, Aydin et al. \cite{aydin2019general} devised a novel multi-fidelity computational framework to train fully connected neural networks and applied it to predict the deformation of a thin elastic membrane. Although the method appears promising for elastic scenarios, it might be less suited for cases with non-linear material behaviour, such as envisaged here. 
The reason is that for high fidelity simulations of heterogeneous and high mechanical contrast materials with inelastic constitutive response high accuracy is needed to cope with non-linear effects such as stress and strain localization, shear banding and onset of yielding.

Cecen et al. \cite{cecen2018material} developed a data-driven approach to efficiently link three-dimensional microstructures to their homogenized properties using convolutional neural networks with improved accuracy in property predictions as well as reduction in the computation time compared to conventional microstructure quantification. Fernandez et al. \cite{Fernandez2020} employed artificial neural networks as a surrogate model to transfer atomistic grain boundary decohesion data to continuum scale  modeling of intergranular fracture in Aluminum. Li et al. \cite{li2019hierarchical} developed a hierarchical neural hybrid method to efficiently compute failure probabilities in high dimensional problems employing the multi-fidelity approach introduced by Aydin et al. \cite{aydin2019general}. They showed that for achieving an accurate estimate of the rare failure probability, a traditional Monte Carlo method needs to solve the equations significantly more frequently than the proposed hierarchical neural hybrid method. In addition, Monte Carlo models are generally numerically not very efficient, due to their discrete event probing and associated generation and comparison of values against random numbers. 

Although numerous works on the use of machine learning in materials science have been published, these are often geared towards predicting an average (homogenized) behaviour of the system, based on large input data sets. In contrast to this type of approach, we focus here instead on resolving the local response of the system. More specific, we devise a general AI-based solver for predicting the local stresses in heterogeneous solids with high mechanical contrast features and non-linear material response.
This solver can be used to replace or augment  conventional numerical approaches like finite element methods. In this work, we demonstrate the ability of machine learning to  calculate  mechanical stress field in complex microstructures with both, elastic and elasto-plastic material behaviour. For the sake of simplicity, isotropic elastic and elastic-perfect-plastic response (zero hardening) are adopted here.
We focus here only on stress, since it appears explicitly in the partial differential equation for mechanical equilibrium. In principle, the same framework could be used to predict strain as well. Furthermore, when knowing the constitute behaviour of the material, in specific cases, it may be possible to calculate strain components from the predicted stress field through post-processing: the elastic stress is through Hooke's law always linearly related to the local stress but the plastic strain is path-dependent and may be more challenging to resolve. In such cases an equivalent machine learning approach as used here for stress can also be applied without any loss in generality.

In \secref{sec_method} a brief review of the governing equations for mechanical equilibrium in elasto-plastic solids is presented followed by an overview of the machine learning algorithm we use in this study. \secref{sec_setup} is dedicated to the simulation setup and the generation of the mechanical stress field data that serve for training and evaluating the machine learning network. The results and their discussion for all trained scenarios with different constitutive response and microstructure topologies are presented in \secref{sec_result}, placing attention on both, the performance of the network on geometries and mechanical contrast ranges similar to the trained ones and also on those beyond the training range. Conclusions and future opportunities are presented in \secref{sec_conclusion}.

\section{Method}
\label{sec_method}
\subsection{Short summary of large deformation elasto-plastic solid mechanics}
Here we present a concise summary of large-deformation elasto-plastic solid mechanics, as implemented in the D\"usseldorf Advanced Material Simulation Kit (DAMASK) \cite{ROTERS2019420}. For a complete description as well as the details of the different numerical implementation, parameter identification and solution algorithms, the reader is referred to the original papers \cite{ROTERS2019420,ROTERS20101152}. Assuming large deformations, neglecting inertial and body forces, the strong form of the mechanical equilibrium in a continuous domain is
\begin{equation}
\begin{array}{ccc}
    \mathrm{Div}\, \bm{P} & = & \bm{0} \;\; \mathrm{in} \;\;  \Omega_{0}  \\
\bm{P} & = & \bm{P}_\mathrm{BC} \;\; \mathrm{on} \;\; \Gamma_{P} \\
    \bm{u} & = & \bm{u}_\mathrm{BC} \;\; \mathrm{on} \;\; \Gamma_{u} \\
\end{array}
\label{EqMechSolid}
\end{equation}
where $\bm{P}$ is the first Piola-Kirchhoff (PK) stress tensor and $\Omega_{0}$ is the volumetric domain. The above partial differential equation (PDE), together with the boundary conditions (prescribed either as traction or displacement on non-overlapping surfaces $\Gamma_{P}$ and $\Gamma_{u}$ of $\Omega_{0}$), describe the strong form of the mechanical equilibrium. Under external or internal loads, the domain will deform and the material points will move from their reference positions $\bm{X}$ to their current positions $\bm{x}$ through the deformation field $\bm{\chi}$. In the context of large-strain solid mechanics, the deformation gradient $\bm{F} = \mathrm{Grad}\, \bm{\chi}$ is multiplicativly decomposed into elastic ($\bm{F}_\mathrm{e}$) and plastic ($\bm{F}_\mathrm{p}$) parts
\begin{equation}
\begin{array}{ccc}
\bm{F} = \bm{F}_\mathrm{e} \bm{F}_\mathrm{p}
\end{array}
\label{EqDecompos}
\end{equation}
with the plastic flow rule 
\begin{equation}
\begin{array}{ccc}
\dot{\bm{F}}_\mathrm{p} = \bm{L}_\mathrm{p} \bm{F}_\mathrm{p}.
\end{array}
\label{Eqflowrule}
\end{equation}

In a first approach we assume that the material undergoes isotropic plastic deformation when loaded beyond its yielding point. This means that the inelastic deformation is governed by the second invariant of the deviatoric stress tensor, referred to as $\mathrm{J}_2$. In other words, tensorial directionality is reduced to a scalar stress value (which is typically calculated through the von Mises equivalent stress measure) to which the strength of the material (plastic resistance against inelastic deformation) is compared. Under the assumption of such an isotropic $\mathrm{J}_2$ plastic material response model, the velocity gradient $\bm{L}_\mathrm{p}$ is formulated as
\begin{equation}
\begin{array}{ccc}
\bm{L}_\mathrm{p} = \dot{\gamma}_\mathrm{p}\dfrac{\bm{S}^\mathrm{dev}}{||\bm{S}^\mathrm{dev} ||}
\end{array}
\label{EqVelGrad}
\end{equation}
with $\bm{S}^\mathrm{dev} = \bm{S} - \mathrm{tr}\, \bm{S} / 3$ the deviatoric part of the second PK stress ($\bm{S}$) and
\begin{equation}
\begin{array}{ccc}
\dot{\gamma}_\mathrm{p} = \dot{\gamma}_\mathrm{0} \Bigg(\dfrac{S_\mathrm{vM}}{S_\mathrm{y}}\Bigg)^n
\end{array}
\label{EqGammDot}
\end{equation}
the plastic strain rate. $\dot{\gamma}_\mathrm{0}$ and $S_\mathrm{y}$ in \equref{EqGammDot} are the initial strain rate and yield stress, respectively. $n$ is the plastic yielding exponent that describes the rate sensitivity and 
\begin{equation}
    S_\mathrm{vM} = \sqrt{\dfrac{3}{2}} ||\bm{S}^\mathrm{dev}||
    \label{eqVM}
\end{equation}
the von Mises stress equivalent measure of $\bm{S}$.
Once the total deformation is decomposed into its plastic and elastic parts through an iterative solution of \equref{EqDecompos}-(\ref{EqGammDot}) as explained in \cite{ROTERS2019420}, the stress is calculated by the generalized Hooke's law
\begin{equation}
\begin{array}{ccc}
\bm{S} = \msbi{C} : \bm{E}
\end{array}
\label{Eq2PK}
\end{equation}
where 
\begin{equation}
\begin{array}{ccc}
\bm{E} = \frac{1}{2}(\bm{F}_\mathrm{e}^{\mathrm{T}}\bm{F}_\mathrm{e} - \bm{I})
\end{array}
\label{EqGreenStrain}
\end{equation}
is the Green strain tensor and $\bm{S}$ is the second PK stress, related to $\bm{P}$ as
\begin{equation}
\begin{array}{ccc}
\bm{P} = \mathrm{det}\bm{F}_\mathrm{p} \bm{F}_\mathrm{e} \bm{S} \bm{F}_\mathrm{p}^{\mathrm{-T}}.
\end{array}
\label{Eq1PK}
\end{equation}
The first PK stress introduced above, should satisfy the mechanical equilibrium in \equref{EqMechSolid}, which is solved numerically in DAMASK with the help of a spectral solver \cite{SHANTHRAJ201531,EISENLOHR201337}. The non-linear systems of equations outlined above are solved in iterations until convergence is achieved. These solutions serve here as a reference for training as well as for assessing the quality and the predictive capability of the machine learning method.

In this work, we investigate two constitutive test cases of purely elastic and elasto-plastic materials with perfect plastic (zero strain hardening) response. For simplicity, only isotropic elasticity is considered here. These constitutive assumptions reduce the material properties to two (three) parameters of  Young's modulus $Y$, Poisson's ratio $\nu$ (and yield stress $S_\mathrm{y}$) in case of elastic (elasto-plastic) material, respectively. The mechanical heterogeneity of the material is then mapped as a topological aggregate (mimicking a polycrystal) where each domain assumes a set of different material parameter values in the ranges of $[60,120]$ GPa, $[0.1,0.4]$ and $[50,200]$ MPa for $Y$, $\nu$ and $S_\mathrm{y}$, respectively.

\subsection{A machine learning approach based on U-Net}
Using machine learning and specifically deep neural networks has become ubiquitous in material science (see Refs.~\cite{sha2020artificial, Ye2018,Butler2018,Schmidt2019, Ramprasad2017,Niezgoda2013,McDowell2016,Bereau2016,wodo_broderick_rajan_2016,Agrawal2016,Dimiduk2018, MASI2021104277} for a review). 
Most of the current ML related innovations in materials science and engineering successfully aim at accelerated material discovery \cite{Raccuglia2016a, Meftahi2020, Sun2019, DAI2020109618}, 
efficient interatomic potential development~\cite{behler2007generalized, Kolb2017,Behler2016,Dragoni2018}
or feature identification in complex pattern that have relevance for materials performance \cite{Jiang2021,Yoshitaka2016,DECOST201730,DECOST2015126,DECOST2017438}.
However, we show here that ML can also help to fundamentally change the way how we solve (non-linear) partial differential equation systems in conjunction with advanced constitutive laws that describe complex material microstructures, much faster than via classical finite element or spectral solvers \cite{RAISSI2019686, Rad2021a}.
This tenet change  has the potential to revolutionize continuum-based simulations of materials, allowing a substantial enhancement in the accessibility of materials and topologies of high complexity, size and speed, to quantitative predictions.

Introducing a deep learning alternative for a physics-based solution scheme should be carried out with stringent assessment of the performance in quality of the solutions (see Refs.~\cite{decost2020scientific, vasudevan2021off} for a review on challenges of introducing AI as a new tool in material science). The necessity of a quantitative quality assessment of the predictions, among other factors, is motivated by a specific common challenge associated with all  deep learning approaches: unlike in  conventional solver  schemes, where the solution is directly built upon the fundamental governing equations (e.g.~\equref{EqMechSolid}), in deep learning, the network only learns to \textit{reproduce} the correct output based the training data. 
The training data-set has the correct value of output which are calculated by the physics-based conventional solver (hence, theory-trained AI~\cite{Rad2021a}), and this establishes an indirect way for the network to be exposed to the physics of the problem. In other words, the neural network learns a mapping between the inputs and their corresponding outputs which are calculated by a physics-based solver. Although the network does not directly incorporate the physical laws, it can mimic the outputs which are based on these physical laws. 
This means that deep learning is for such tasks well equipped for  digesting and reproducing  pattern features very efficiently and relate patterns to topologies, but this  does not per se include gaining insights into the underlying physics-related origin of certain features of such patterns (except for the topological ones). Thus, when using deep learning  primarily as an efficient solver rather than as a physics  tool seems to exploit its biggest strength in the current context.
This lack of direct inclusion of the physical laws (and in general human knowledge) motivated invention of procedures to shed light on how the neural networks make their predictions; these methods are commonly referred to as  explainable AI methods (for example see Refs.~\cite{doran2017does, miller2017explainable, preece2018stakeholders, arrieta2020explainable}). In this article, unlike the common approaches in explainable AI (e.g. Ref.~\cite{bach2015pixel, arras2017relevant}), we develop an understanding about the relationship between the inner structure of the proposed neural network and its capability in finding an approximate solution surrogate to the posed boundary value problem by using a number of tests. Utilizing these findings, we make connections between details of the machine learning solution to the well defined solution algorithms used for solving PDE with conventional spectral solvers.

Neural networks vary in their basic neural units, the arrangement of these units in the layers and their connectivity, the character of the loss functions (e.g in terms of the quantification of the deviations of the predicted values relative to the reference values in the training data), and the 'reductionist' spirit of the network design (e.g. see Refs.~\cite{hochreiter1997long, ronneberger2015u, lechner2020neural}).
This diversity in possible architecture details is reflective of the diversity of the applications. For applications with new characteristics (and constraints) one might require to develop a network topology and network workflow that matches the problem solving at hand more adequately. Examples of such novel architectures which have been tailor-made for specific physics- and materials science applications can be found in  Refs.~\cite{linka2020constitutive, mohan2020embedding}.

In this article, our goal is to develop and to use a neural network architecture which is capable of estimating the local stress fields in heterogeneous media exposed to external loads, using only the local material properties and the microstructure topology (e.g. the domain, grain or phase structure) as input information. This requires the network to be able to (i) capture local features efficiently, and also (ii) have the same dimensions for inputs and outputs. These are also requirements for many computer vision problems like segmentation and object detection. Therefore, as a first step, we implement one of the most common neural network architectures in computer vision which is known as the "U-net"~\cite{ronneberger2015u}. U-net was originally developed for biological image segmentation and has since been one of the most commonly used architectures in solving computer vision problems.

Another consideration, which favors using the U-net architecture, arises from the main equation to be solved, i.e. \equref{EqMechSolid}. The operator in this equation is a derivative, which could be captured effectively using the convolutional layers of the U-net. With all the above mentioned considerations, we find adapting this architecture particularly promising for the application of solving the mechanical equilibrium and local material response in solid mechanics. 

\section{Setup and Training}
\label{sec_setup}
\subsection{Generating simulation data}
We used Voronoi tessellation to create 1000 two-dimensional random geometries with 20 different domains in each simulation box. Note that the difference between each domain is not the crystallographic orientation in our current training set-up, but the elastic and plastic material properties are different in each domain. The size of the simulation box is $256\times256\times1$ grid points. Two cases are considered, one with isotropic elasticity and a second one with elastic-ideal-plastic response (i.e. no strain hardening). The material properties of each domain are assigned randomly from the sets listed in Table \ref{tabMatprop}.

\begin{table}[H]
    \centering
    \small
    \begin{tabular}{|c|c|c|}
    \hline
             & Elastic & Elasto-plastic  \\ \hline
$Y$ (GPa)    & $\{80,90,100, 110, 120\}$ & $\{60, 80, 100, 120 \}$ \\ \hline
$v$          &  $\{0.1, 0.2, 0.3, 0.4\}$       &  $\{0.1, 0.2, 0.3, 0.4\}$ \\ \hline
$S_\mathrm{y}$ (MPa)    &     -    &      $\{50, 100, 150, 200\}$       \\ \hline
    \end{tabular}
    \caption{Sets of the material properties used in generating the training data.}
    \label{tabMatprop}
\end{table}

A random choice of Young's modulus $Y$ and Poisson's ratio $\nu$ from Table \ref{tabMatprop} (elastic column) is assigned to the domains (which mimic grains) in the 1000 different geometries that were generated for conducting the full-field simulations in DAMASK. The DAMASK simulations then serve as training data for the U-net. For the case of elasto-plastic material response, in addition to $Y$ and $\nu$ taken from the elasto-plastic column in Table \ref{tabMatprop}, we assigned different yield stress values, $S_\mathrm{y}$, to each domain of the simulation box as well. Furthermore, in case of elasto-plastic material response, $n=20$ and $\dot{\gamma}_0=10^{-3}$ 1/s are adopted as parameter values in the plastic strain rate equation, i.e. \equref{EqGammDot}. These parameters result in a perfect plastic model (zero strain hardening) with different yield stresses in  adjacent domains (i.e. grains).

\begin{figure}[ht]
    \centering
    \includegraphics[width=0.7\textwidth]{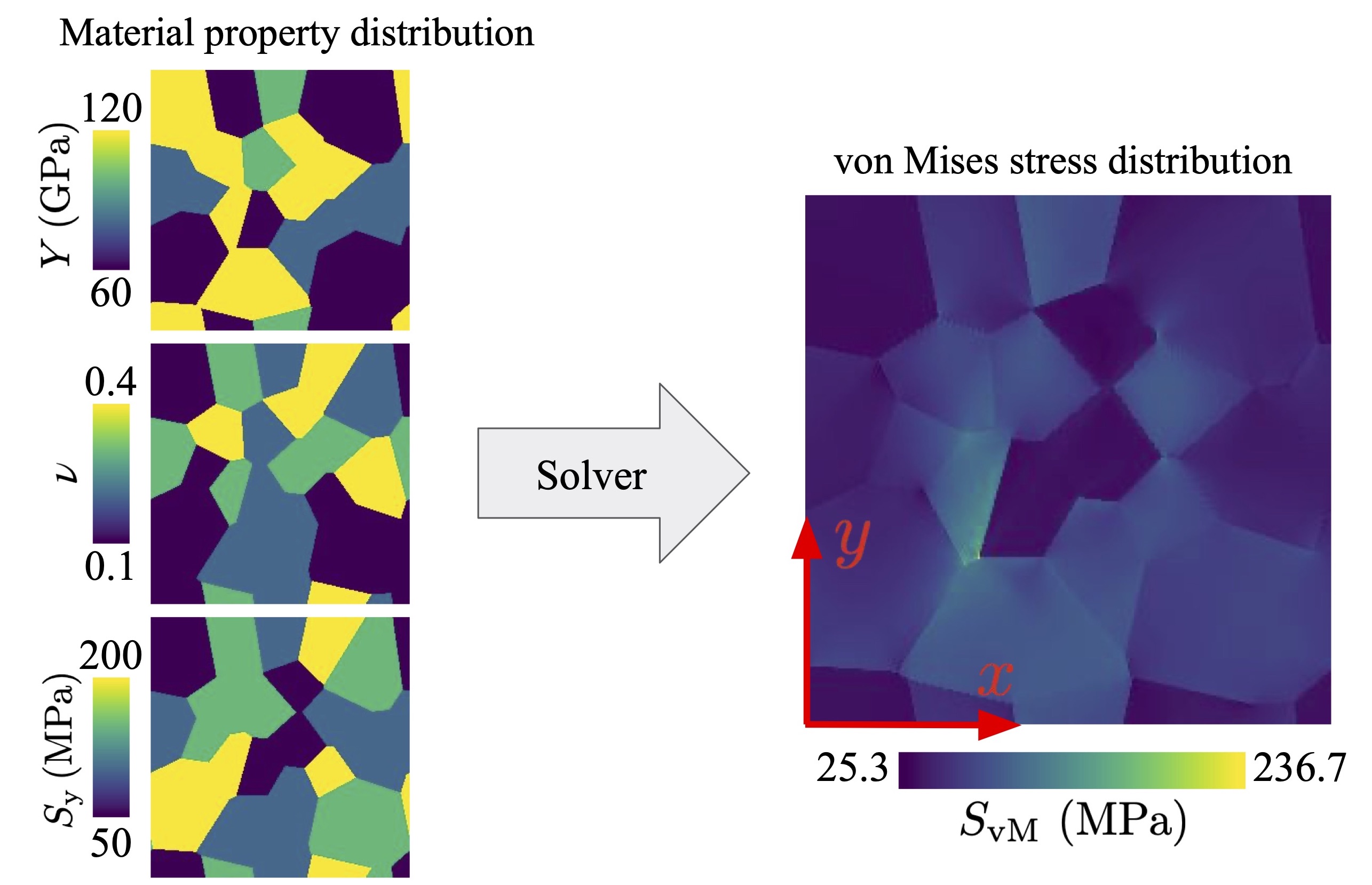}
    \caption{Example of a typical set of maps of an input material property distribution (left) and corresponding output in terms of von Mises equivalent stress distribution (right), calculated with the open-source micromechanical simulation software package DAMASK \cite{ROTERS2019420}. The uniaxial tensile load is applied in $x$ direction. The full-field forward simulations serve as training data for the U-net-based machine learning solver.}
    \label{fig:InOutSchem}
\end{figure}

Once the geometry and the material properties have been assigned, all simulation boxes (i.e. the polycrystal aggregates) were loaded to a tensile strain of $E_{xx} = 0.01$ or $E_{xx} = 0.001$ for elastic or elasto-plastic cases, respectively (see \figref{fig:InOutSchem} for coordinates). The elasto-plastic load was set to a modest value that fell in the middle of the range of the yield strength values in order to enhance complexity, i.e. to obtain a mixture of domains in elastic and elasto-plastic regimes. Stress components in all other normal directions (i.e ${yy}$ and ${zz}$) as well as deformation in the off-diagonal components were set to zero. In case of elastic material behaviour, the load was applied instantaneously in one step, while in the elasto-plastic case, it was applied sequentially over 100 steps. The calculations were performed using the open-source micromechanical simulation software package DAMASK \cite{ROTERS2019420}. Note that even in case of the linear elastic model, the PDEs are not linear due to the non-linear nature of the underlying strain model presented in \equref{EqGreenStrain} that is capable of modeling both large deformations and rigid rotations.

The inputs to the machine learning algorithm are the distributions of the material properties ($Y$, $\nu$ in case of elastic and $Y$, $\nu$, $S_\mathrm{y}$ in case of elasto-plastic response, respectively), represented as color coded images. The output or respectively target of the ML training lies in the prediction of the von Mises equivalent values of the second PK stress, $S_\mathrm{vM}$ as defined in \equref{eqVM}, which is plotted at the end of each simulation as a color coded image. Note that each grid point in the simulation is represented by one pixel in the input/output images (i.e. all images are of size $256\times256$). An example of a set of such input and output images are shown in \figref{fig:InOutSchem}.

For better visualization for readers, the simulated maps are in this paper color coded using the viridis or seismic color maps. However, note that the neural network is trained and working with monochrome images for each of the scalar fields.

\subsection{Network architecture}

\begin{figure}[ht]
    \centering
    \includegraphics[width=0.99\textwidth]{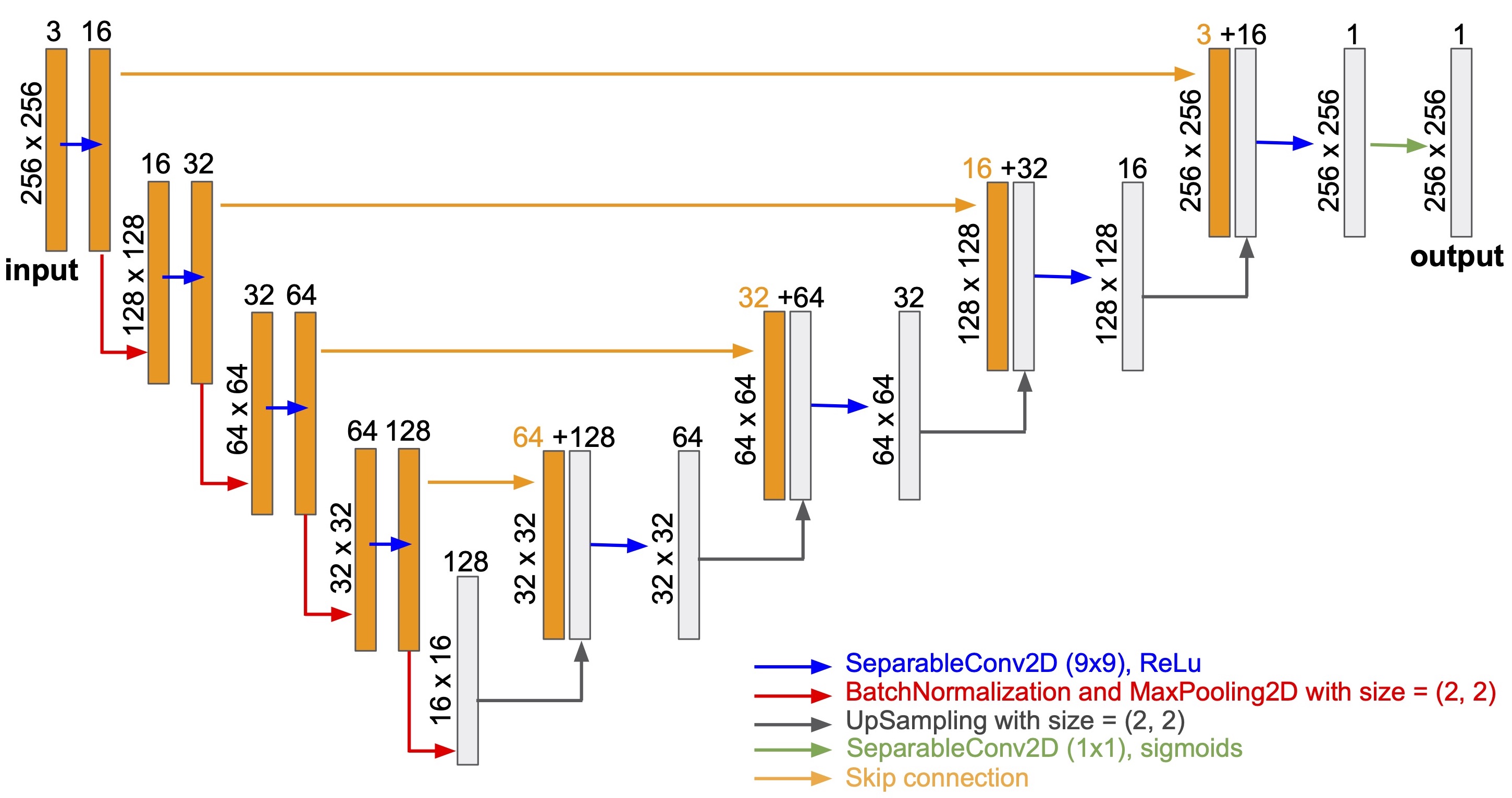}
    \caption{Architecture of the U-net used in this work. Here, the notation of Tensorflow~\cite{tensorflow2015-whitepaper} is adopted for naming the layers. The layers consist of separable convolutional layers (SeparableConv2D) with either a rectified linear unit (ReLu) or sigmoid activation functions to extract the features and apply the non-linearities, batch normalization (BatchNormalization) to transform the layers' outputs to a mean value of zero and a standard deviation of 1, max pooling  (MaxPooling2D) for coarse-graining, and up sampling (UpSampling2D) for going from the coarse grained image to a high resolution one. The skip connections send the images from each contracting step to its expanding counterpart.}
    \label{fig:U-net-original}
\end{figure}

The network architecture used here is an adaption of the U-net~\cite{ronneberger2015u}, see \figref{fig:U-net-original}. The network has two distinct parts, a contracting and an expansive part. The contracting part consists of $N_s=4$ steps where each step is a repeated stacking of contracting modules where each module consists of a convolutional layer, a non-linear activation function, and a batch normalization layer, which is followed by a downsampling layer. The convolutional layer increases the number of channels and the downsampling decreases the number of rows and columns of the data (i.e the width and height of the image). After these $N_s$ steps conducted in the contracting part, the expansive part of the network starts, which also consists of $N_s$ steps. Every step in the expansive part is similar to the contracting part except that in this case the convolutional layers decrease the number of channels and the downsampling is replaced by upsampling which increases the number of rows and columns. Following the U-net architecture, in each downsampling step, the width and the height of the image are reduced by a factor of two. Corresponding to each downsampling step, an upsampling step is implemented where the scaling factor is again two. As the number of downsampling steps equals that of the upsampling steps, the output data have the same dimensions as the input. This architecture also enables us to pass the information of each step of the downsampling sequence directly to the corresponding upsampling part, as it is done in the original version of U-net as well~\cite{ronneberger2015u}. The last contraction step is followed by a convolutional layer which reduces the number of channels to 1 and applies a sigmoid function which maps the values into the range [0, 1].
In spite of its overall similarity to the original U-net structure, the variant of the neural network that we use here deviates from the original design, as explained in the following.

First and foremost, the kernel size of the convolutional layer in our model ($k=9$) is much larger than the kernel size used regularly in U-net (i.e. $k=3$). Here, a kernel of size $k$ refers to a $k \times k$ matrix whose elements are to be learned during the training of the network. The kernel, i.e. specifically its matrix size, is a key feature of convolutional neural networks that is applied step-wise in the form of a sliding array across pattern data.
This process has the aim to amplify, identify and extract certain features from an input image. It is usually coded in the form of a simple matrix (of much lower rank compared to the image size), that is sequentially slid across the image and multiplied at each sequential position with a subset of the input array such that the output enhances certain topological pattern features such as edges, corners, gradients, etc. The rationale behind our choice for increasing the kernel size from 3 to 9 is that a larger kernel size can lead to a more accurate derivative estimate. In fact, as shown in the  results section, we observe that for small kernel sizes the network's capability to predict the results is significantly reduced. The analysis shows that the kernel matrix size is indeed an essential parameter for rendering this approach a viable solver alternative, capable for instance of picking up local stress peak and stress gradient features. These local stress features are related to the mechanical contrast variations which are characteristic of  heterogeneous  materials. 

Another important modification which we implemented in our version of the U-net lies in  replacing the conventional convolutional layers by separable convolutional layers. In contrast to the convolutional layer used in the original U-net design, where the outputs obtained by 
applying the kernel to each channel are summed up for all the channels, in our new approach  the outputs are added together with weights that are learned. Using an unweighted sum is a more suitable choice for segmentation applications as one could expect high correlation between different channels of the input, e.g. between the red, green, and blue colors of an image. For our application, however, we are not expecting such a high correlation between the different material properties (i.e. between the Young's modulus, Poisson's ratio, and the yield stress). The depth of the U-net (i.e. the number of contraction/expansion steps) and also the number of channels are modified in our implementation. Effects that stem from an increase in the depth of the U-net are studied systematically and discussed in the results section. 

\subsection{The input data: random microstructures with isotropic constituent parameters}
The input data include in the present case the spatial distribution of the material properties, including the local Young's modulus $Y$, the local Poisson ratio $\nu$, and in the case of the elasto-plastic behavior, the local yield strength values $S_\mathrm{y}$. We arrange this information by stacking  $(w, h)$ arrays of $Y$ and $\nu$ to form a $(w, h, 2)$ array, with $w=256$ and $h=256$. For the elasto-plastic problems, an extra channel is added which contains the yield stress values $S_\mathrm{y}$, as an isotropic measure of the material's resistance against inelastic shape change. We also introduce an additional channel for $Y/\nu$. The choice of this additional feature is motivated by the common form of the solutions emerging from elasticity theory, where $Y$ and $\nu$ appear in mixed terms of $Y/\nu$ or $Y/(1+\nu)$. The values in the channels are shifted and scaled such that they all  fall into the range $[0, 255]$. 

\subsection{The output data: von Mises stress distribution in heterogeneous media}
Using the input data, i.e. the scaled material properties, we aim to predict the distribution of the von Mises stress introduced in \equref{eqVM}, again in  scaled form. The von Mises stress value is an equivalent stress measure which reduces a tensorial form to a scalar surrogate for isotropic cases.
In general, we are here focusing on the von Mises stress measure as it is a first order parameter that allows to reveal the most important key features of the mechanical heterogeneity in isotropic media subjected to elastic-plastic loads.

Note that in our AI-approach, we do not calculate the von Mises stress by separate prediction of the individual stress components, but we calculate it directly in an end-to-end approach, i.e. using the inputs mentioned above. We speculate that calculation of the individual stress tensor components would be more straightforward for the neural network, but this would require the use of more output channels. For highly mechanically anisotropic cases it would be however pertinent to calculate the components of the stress tensor individually (and extract from these also secondary measures such as equivalent stress values).

\subsection{Training the network}
The neural network is implemented, trained, and tested using Keras~\cite{chollet2015keras} which is an application programming interface (API) written in Python, running on top of the machine learning platform TensorFlow~ \cite{tensorflow2015-whitepaper}. The 1000 samples are split randomly into the training and test sets of size 950 and 50 samples, respectively. To train the network, the mean absolute error is used as the loss function (which should be minimized). The minimization of the loss function is done using the ADAM optimizer which is a stochastic first-order gradient-based method \cite{kingma2014adam}. We use random samples in batches of size 32 for the gradient estimation, and continue training for 400 epochs, i.e. 400 complete passes through the whole training dataset. We set the hyperparamters of the ADAM optimizer to $\beta_1=0.9$, $\beta_2=0.999$, $\epsilon=10^{-7}$, and a learning rate within the range $[1,5] \times 10^{5}$. The training is carried out for 400 to 800 epochs. During the training the mean absolute error of $\simeq 0.02$ is achieved without any significant sign of over-fitting.

\section{Results and discussion}
\label{sec_result}

\subsection{Isotropic elastic case}
Once the network is trained using the training dataset outlined above, we evaluate it with sets of input material property distributions in topological configurations and parameter combinations which had not been used for training the network. More specific, the evaluation on the test dataset is done both (i) for geometries similar to the training data (i.e. Voronoi tessellations), and (ii) geometries which are completely different from those that had been used for training. For the geometries in (ii), we included cases where the surface of the inclusion (i.e. the internal domain) is curved, as a test for the model's capability to generalize beyond Voronoi-based representative volume elements, where straight, faceted interface shapes were used. These results are presented in the following sub-sections. 

\subsubsection{U-net predictions for materials with domain topologies similar to those of the training data}
\label{sec_simgeom}

\begin{figure}[ht]
\centering
    \includegraphics[width=0.7\textwidth]{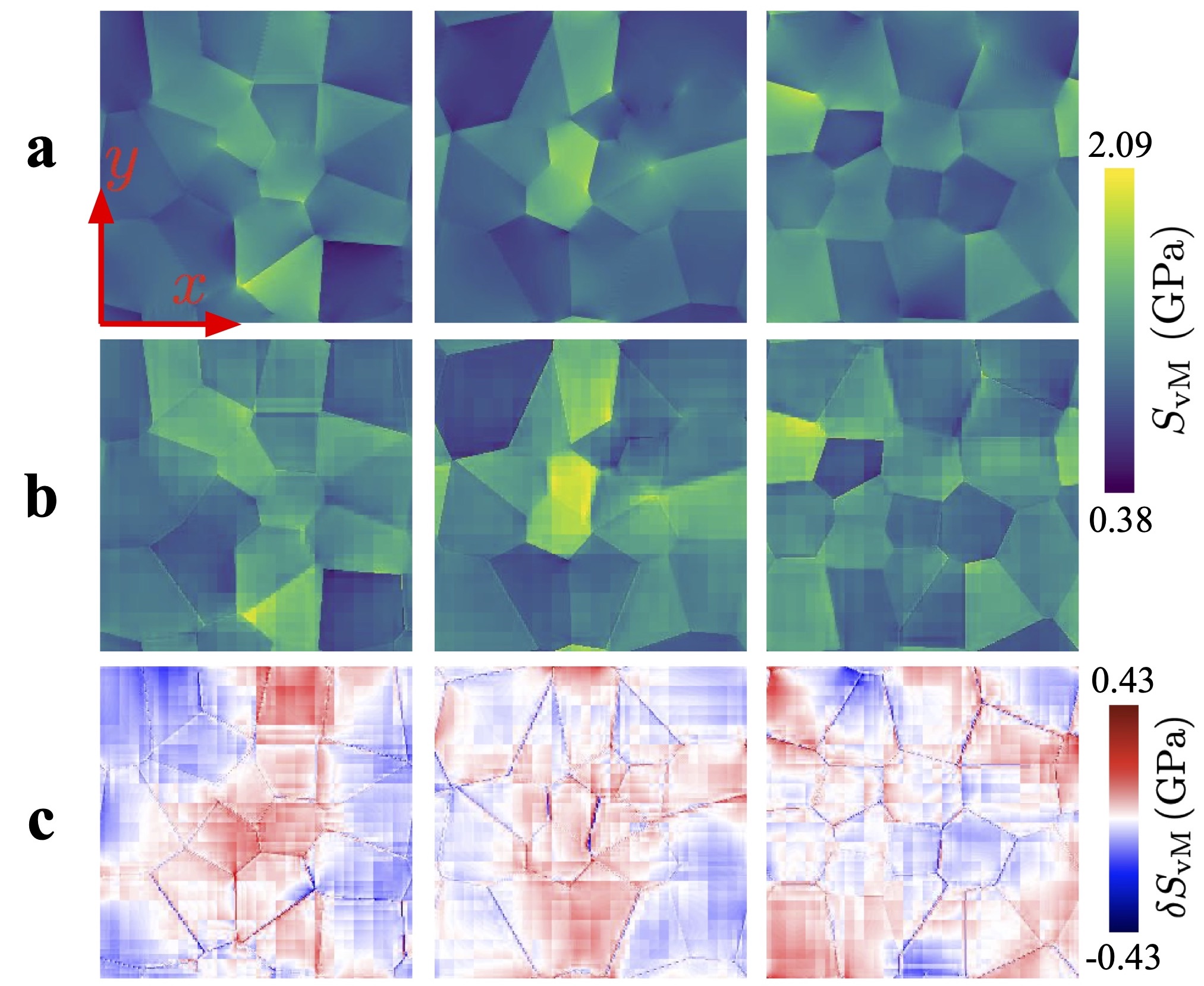}
    \caption{Von Mises equivalent stress simulation results obtained by DAMASK in conjunction with a spectral solver \textbf{(a)} and the corresponding AI predicted \textbf{(b)} von Mises stress distribution for three Voronoi tessellations of test microstructures (not included in the training dataset) using the modified U-net approach. The error, calculated as deviation between the two von Mises stress distributions, $\delta S_\mathrm{vM}$, is shown in \textbf{(c)}. }
    \label{fig:Voronoi_Pred}
\end{figure}

Von Mises stress predictions based on the U-net architecture for three Voronoi tessellations from the test data set are shown in \figref{fig:Voronoi_Pred}.
The figure reveals that the AI-based solution, obtained by the modified U-net method, qualitatively and in part even quantitatively agrees well with the DAMASK solution, although there are regions with a higher relative error of up to 25\%. As  will be shown in the following, these regions of high deviation are limited to only a few rare points and for most of the grid points in the domain the relative error is below 4\%. In particular, the surrogate model is able to adequately reproduce the sites of stress concentrations and the distribution at the domain junctions, which are due to local contrast in elastic constants. 

\begin{figure}[ht]
    \centering
    \includegraphics[width=0.9\textwidth]{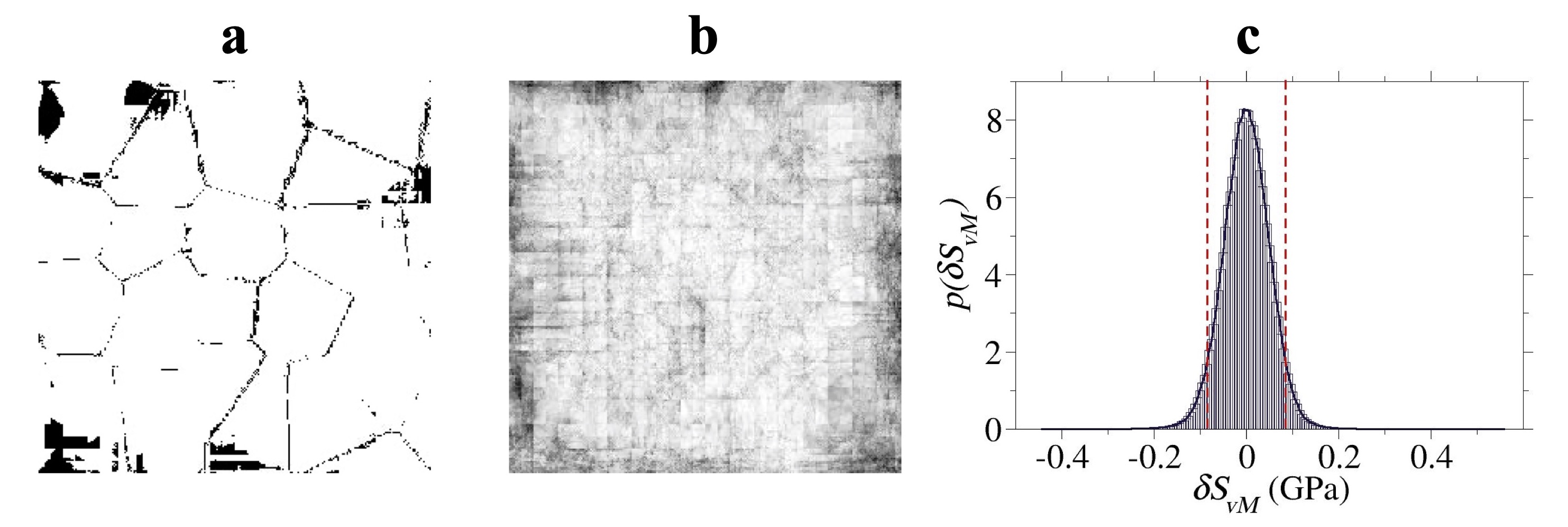}
    \caption{Distribution of the absolute values of the deviations between the reference data obtained from the full-field DAMASK based spectral solver simulations and the U-net predictions \textbf{(a)}, showing only regions with errors above 83 MPa (4\% relative error) in  black color for one of the test cases. \textbf{(b)} The local probability of occurrence of errors larger than 83 MPa. The probability is shown by a gray-scale color-code (the white and black colors represent the probability of zero and one, respectively). This quantity is calculated by averaging the observations similar to (a) over all the test data-sets. In this procedure for each of the test samples, a value of 1 is assigned to regions with an error above  83 MPa, and 0  otherwise. \textbf{(c)} Error distribution for all points in all of the 50 cases in the test data-set with 4\% relative
    error indicated as red dashed lines. }
    \label{fig:Error}
\end{figure}

To further investigate the source of the errors reported above, we re-plot \figref{fig:Voronoi_Pred} (c) right, in \figref{fig:Error} (a), showing only regions with error values higher than 83 MPa (4\% relative error) as black. This analysis, firstly, suggests that  there are localized regions of higher error between the domain boundaries where the material properties are discontinuous. This is expected since those regions are typically also the most challenging ones  for  conventional solvers to cope with as well. A possible mitigation strategy for such zones of high local deviation might be the use of filtering schemes in the network, quite similar as it is done in spectral solvers, where the underlying Fourier series can create non-physical high frequency noise in solutions. 

As observed in \figref{fig:Error} (a), in addition to the domain boundaries, there are patches of high error values seemingly close to the boundary of the calculation box. To confirm this, an averaging operation is performed, by simply  adding images similar to the one shown in \figref{fig:Error} (a) of all cases in the test dataset and diving the sum by the number of cases (50 here). The resulting  average error distribution map is shown in \ref{fig:Error} (b). It indicates that these patches are indeed mostly located near the boundaries of the box. We speculate that this is due to the fact that our convolutional neural network, unlike the DAMASK solver, does not consider periodic boundary conditions. More specifically, when the kernel is applied to points that are located close to the outer borders of the simulation box, for the pixels beyond these borders  their periodic counterpart values from the opposite borders should be used, like used in most similar set-ups where classical periodic boundary conditions are enforced. It is therefore likely that enforcing the periodicity of the input images in the neural network will help reducing the boundary effect errors. These aspects refer to work in progress which will be analyzed in more detail and be reported in ensuing publications. A positive aspect of this error close to the borders of the simulation domains, however, is that it seems to be merely a geometrical effect related to the simulation box and not a fundamental problem of the U-net approach to identify surrogate solutions. \figref{fig:Error} (c) shows the error distribution of all points in all cases of the test dataset. The analysis shows that most of the points {(91\% of all points)} fall inside the red dashed lines indicating equal or less than 4\% relative error.  

\subsubsection{Application of the U-net to geometries far from the training data}
\label{sec_fargeom}

\begin{figure}[ht]
    \centering
    \includegraphics[width=0.8\textwidth]{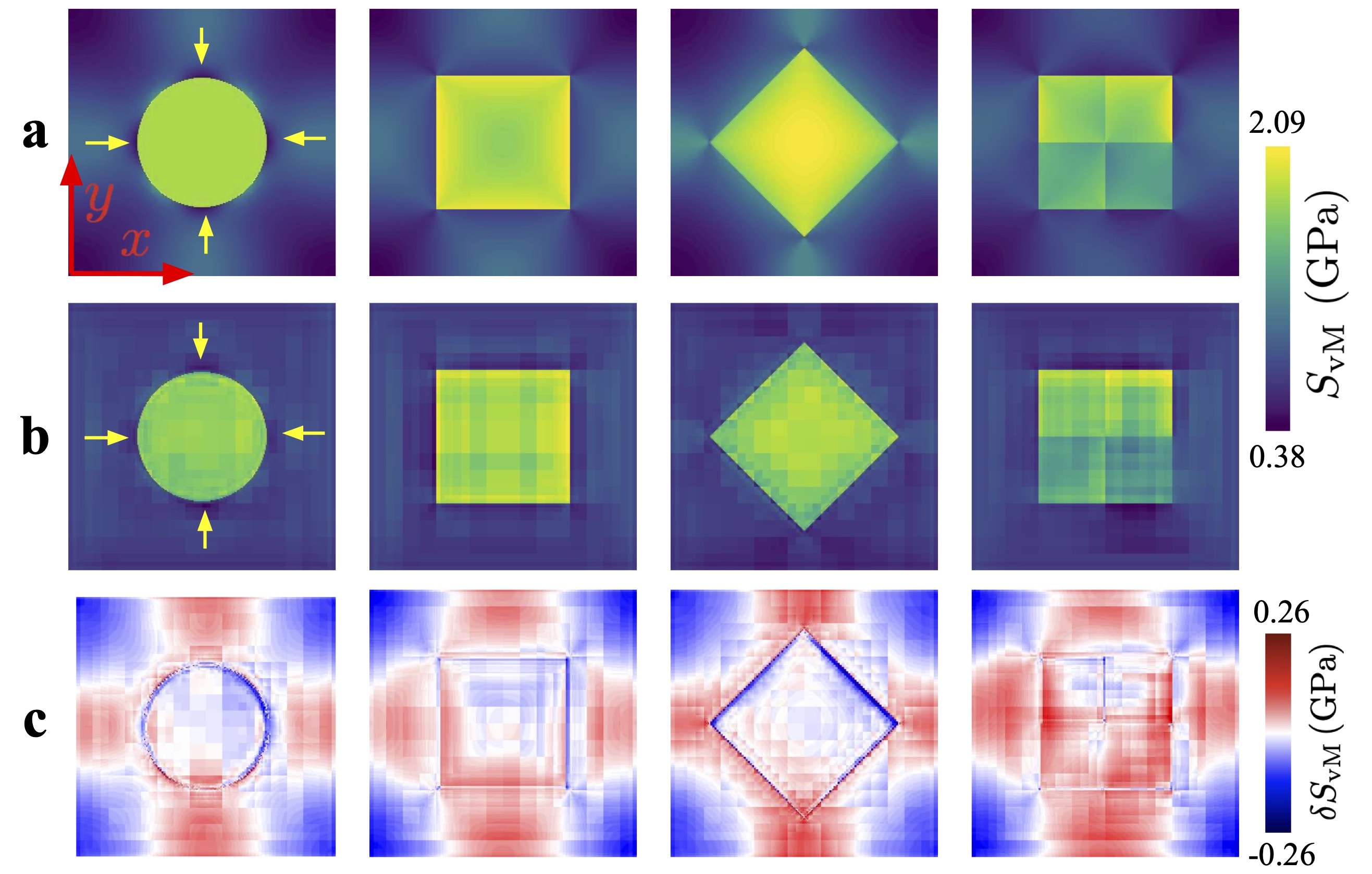}
    \caption{Simulation results \textbf{(a)} for isotropic elastic material response for topological cases far away from the trained set of Voronoi structures and the corresponding AI predicted \textbf{(b)} von Mises stress distribution for different hard inclusions in a soft matrix. The error between the AI prediction and the  DAMASK solution is shown in \textbf{(c)}. }
    \label{fig:SpecialCaseElastic}
\end{figure}

Next, we look into geometries that could not be represented by Voronoi tessellation. These include classical elastic problems of circular and square shaped inclusions with larger Young's modulus embedded in a softer matrix. The von Mises stress distributions for such cases predicted by the AI and calculated by DAMASK are shown in \figref{fig:SpecialCaseElastic}.

Although the topologies of these high-mechanical contrast test cases are fundamentally different from the Voronoi structures that were used for training, the AI-based predictions still capture the correct stress partitioning between the soft and hard phases as shown in \figref{fig:SpecialCaseElastic}. The error between the reference patterns obtained by the full-field DAMASK simulations and the AI-predictions is shown in the third row. The maximum relative error is about 12\%, however, as in the previous cases, the regions of largest error are limited to the domain boundaries. Otherwise, the stress distribution around and inside the hard inclusions follows that observed in the reference pattern. For example, note that for the four deep blue regions around the circular inclusion at 0, 90, 180 and 270$^{\circ}$, as marked by yellow arrows, the minimum von Mises stress  based on the DAMASK-calculated and AI-predicted values are 802.5 and 849.4 MPa, respectively. 

\subsubsection{Effects of depth and kernel size of the U-net}
The Gibbs phenomenon is a classical numerical issue associated with solution algorithms of 
mechanical problems with sharp transition features in material properties (as shown in many works, e.g. \cite{Willot2015,Khorrami2021}), especially when using the fast Fourier transform method as a solver. This undesired oscillating and overshooting effect is produced by Fourier solvers applied to piece-wise continuously differentiable functions at discontinuities (such as here when crossing interfaces). The oscillations in the solution fields due to such sharp transitions can be minimized using multiple filtering schemes \cite{Willot2015}. Next we therefore study in more detail the predicted stress profile and look into the effect of the depth of the network (i.e. the number of down- and upsampling steps) on such oscillations. 
\begin{figure}[ht]
    \centering
    \includegraphics[width=0.85\textwidth]{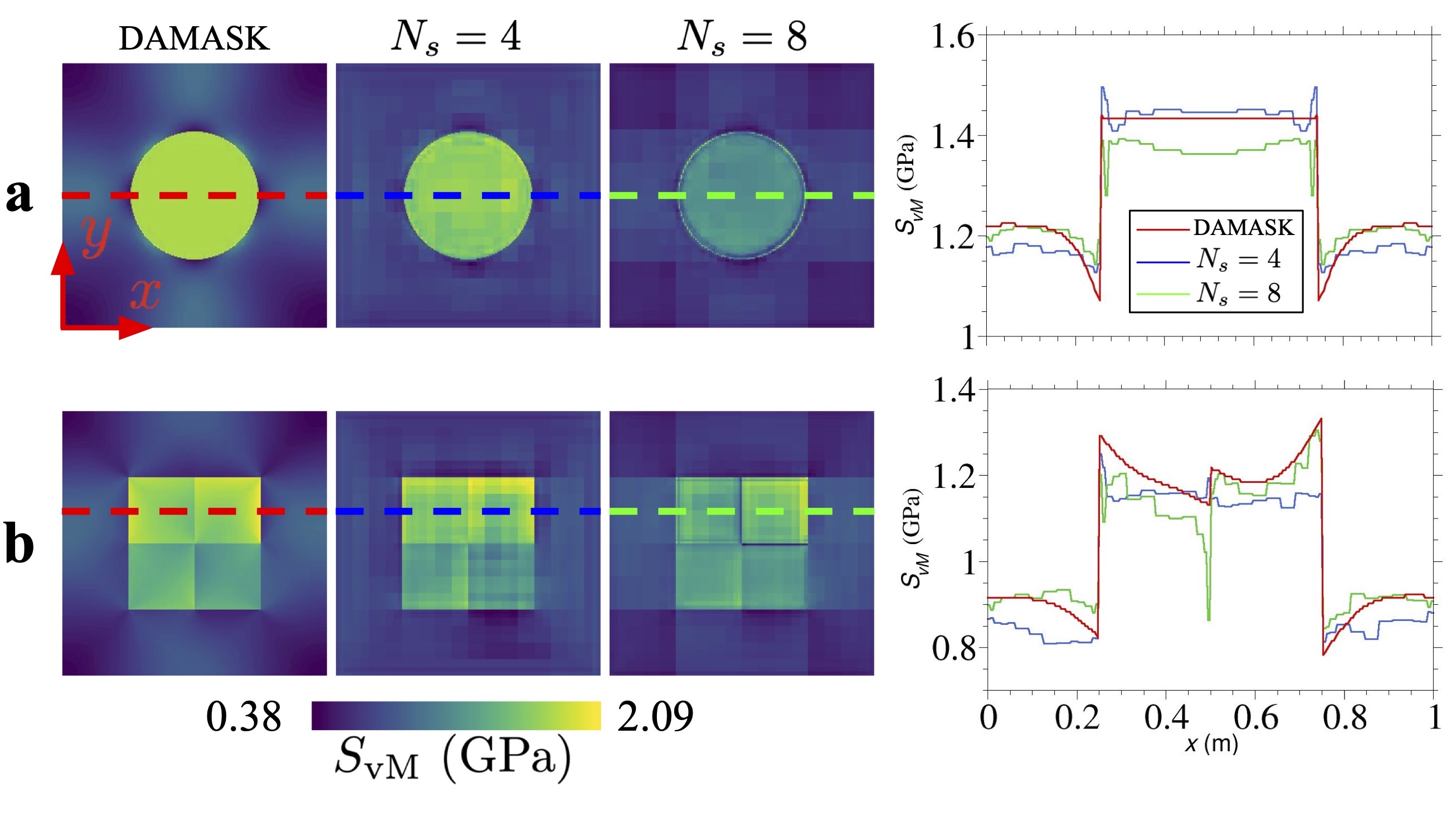}
    \caption{Von Mises stress distribution around circular \textbf{(a)} and square \textbf{(b)} inclusions calculated by DAMASK as well as two network architectures with depths of $N_s=4$ and $N_s=8$. The stress profiles are plotted along the dashed lines with red, blue and green corresponding to DAMASK, network with $N_s=4$ and $N_s=8$ based results, respectively.}
    \label{fig:Depth}
\end{figure}
As shown in Figs.~\ref{fig:Depth}-right column, the deeper network, i.e. the network with $N_s=8$ is capturing the wake of the stress field better than the shallower network with $N_s=4$, however, the jump in the stress magnitude at the transition of the material property is larger. This seems analogous to the full Fourier approach used in spectral solvers where all frequencies are included (deep network) compared to the filtered approach (shallow network). In the spectral method, filtering the high frequencies is equivalent to approximating the derivatives with lower order finite differences (such as forward or backward differences). Since the convolution kernels used in the network could correspond to taking derivatives of the data, increasing the depth of the network, at least conceptually, will be analogous to work with higher order derivatives in conventional solvers, amplifying the fluctuations.

\begin{figure}[ht]
    \centering
    \includegraphics[width=0.8\textwidth]{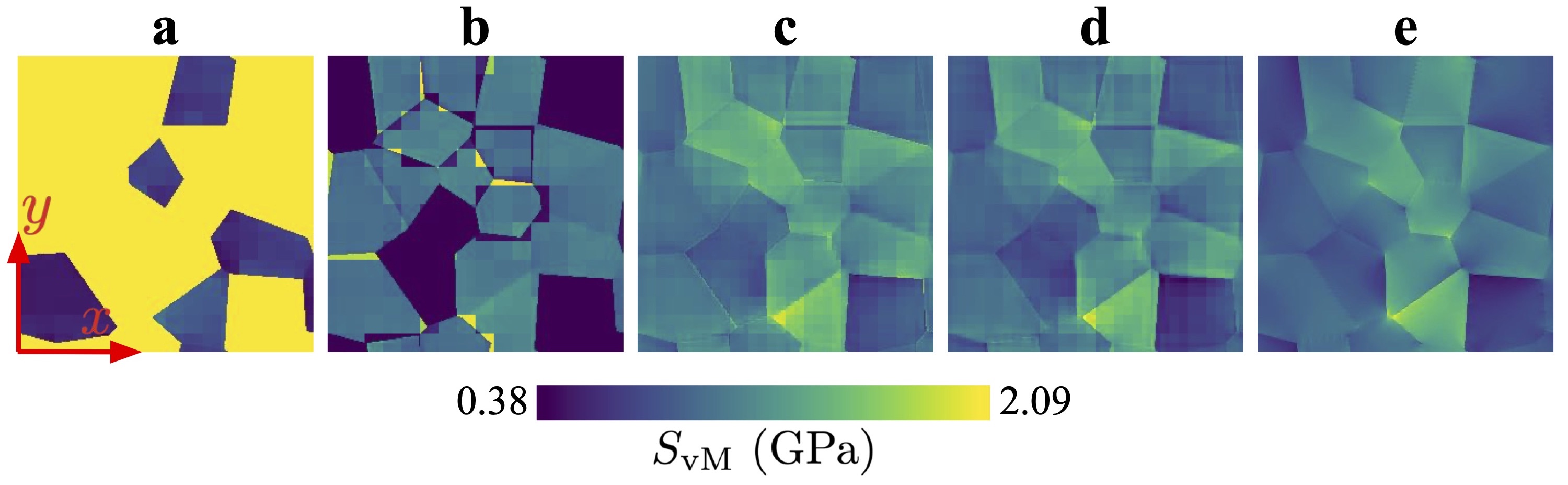}
    \caption{Effect of the size of the $k \times k$ kernels  with $k= 3, 5, 7, 9$ on the predicted von Mises equivalent stress distribution, shown in \textbf{(a)}, \textbf{(b)}, \textbf{(c)}, and \textbf{(d)}, respectively, compared to the results obtained from the full-field DAMASK simulations used here as reference \textbf{(e)}. All models were trained under the same conditions.}
    \label{fig:Kernel}
\end{figure}

Next we focus on the effect of the kernel size on the AI-predicted results. 
As seen in \figref{fig:Kernel}, smaller $k \times k$ kernels with $k=3$ and $k=5$ result in completely incorrect stress distributions. In fact, under the training conditions discussed above, it was not possible to reach the loss values corresponding to networks with larger k values ($\simeq$0.02). However, networks with $k \times k$ kernels of $k=7$ and $k=9$ were trained to the lower loss values and they perform much better. Note that the results discussed in the sections \ref{sec_simgeom} and \ref{sec_fargeom} were based on $k=9$.

\subsection{U-net AI predictions for cases with elasto-plastic material response}
As the next step to create a surrogate AI stress solver, we extend the U-net solution to elasto-plastic problems. It is worth mentioning that in order to obtain the elasto-plastic response in the full-field DAMASK simulations used here as reference, the load must be applied incrementally (here we used 100 steps). This renders the conventional DAMASK simulation procedure computationally more demanding compared to the elastic problems where the load is applied instantaneously. In our AI approach, we can obviously omit the extra computational costs that are caused by this incremental approach. Again, similar to the elastic case, the AI-prediction is, once the system is trained, a single step forward calculation, i.e. one set of inputs, here $(Y, \nu, S_\mathrm{y})$, are fed to the neural network and the von Mises stress is directly predicted, without any iterations. As we have in the current elastic-plastic material case more independent inputs and also a more complicated relation between the inputs and outputs, we first need to increase the predictive power of the neural network. Therefore, we increased the complexity of the neural network by doubling the number of the channels at each stage of the U-net. Based on our observation of the effect of an increase in the depth of the U-net, the increase in the number of channels appeared to be a suited approach for enhancing the complexity of the U-net for predicting mechanical response in heterogeneous media.

Similar as done for  the elastic case, we trained  the U-net using the DAMASK stress predictions to obtain reference values. Due to the nature of the material's elasto-plastic behaviour, the absolute value of the overall applied strain is important. For very small strains, all the domains remain elastic, and for very large strains all the domains enter the elastic-plastic regime. In the former case, we are actually solving a purely elastic problem and consequently the solution depends only on the elastic constants, and for the latter case the stresses could be estimated fairly well from the yield stress values and therefore the elastic constants do not play a significant role. To provide a generic training data-set for the U-net, we choose the magnitude of the applied strain in a range so that some of the domains become plastic while the others remain elastic. In this regime, the stress is determined by a complicated interplay between elasticity and plasticity.

\begin{figure}[ht]
    \centering
  \includegraphics[width=0.99\textwidth]{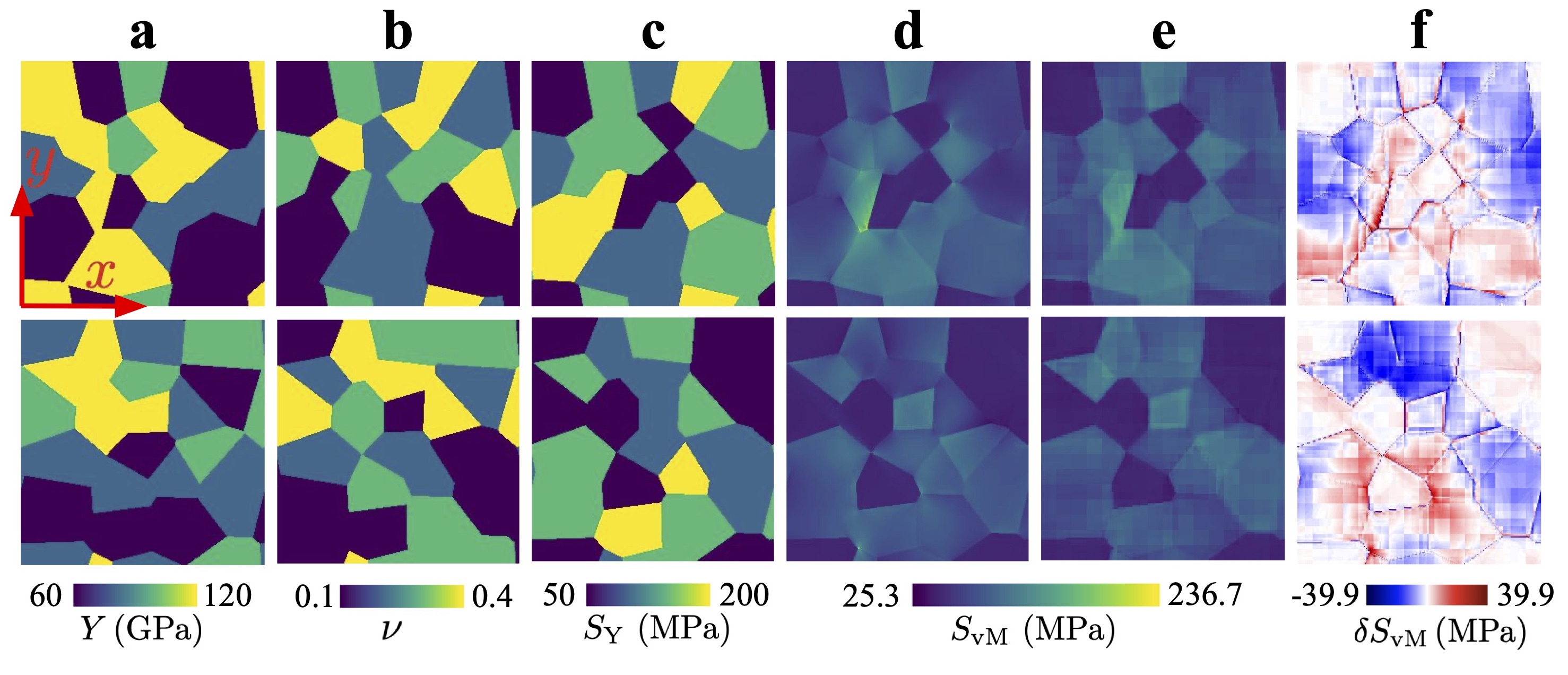}
    \caption{Comparison for two different topologies obtained by  U-net predictions  \textbf{(e)} of the von Mises stress spatial distribution and the corresponding DAMASK calculations \textbf{(d)} for the shown input material property distributions of Young's modulus \textbf{(a)}, Poisson ratio \textbf{(b)} and Yield stress \textbf{(c)}. The error between U-net and DAMASK predictions is shown in  \textbf{(f)}. The applied strain is chosen such that the von Mises stress depends on both elastic constants and the yield stresses, i.e. the spatial pattern of the von Mises stress is neither similar to the elastic constant distribution nor to the yield stress distribution.}
    \label{fig:elastoplastic}
\end{figure}

A typical example of the stress prediction using the trained U-net for the case of elasto-plastic material behaviour is depicted in \figref{fig:elastoplastic}.

\subsection{Speed-up and efficiency evaluation}
In the previous sections, the network training and evaluation were performed on a machine powered by an Intel Xeon CPU clocked at 2.30GHz with a NVIDIA Tesla P4 GPU. However to benchmark the performance of the AI-based solution in comparison with DAMASK, we restrict here both to use only one core of an AMD EPYC 7702 64-Core Processor, clocked at 3343.801 MHz. Limiting the overhead of both calculations as much as possible and running both on the same CPU for a large number of evaluations provides a good measure of the total number of floating point (FL) operations necessary to reach the final result. Of course, working with GPU for AI evaluation could lead to better scalability compared to parallel processing conducted in the typical spectral simulations in DAMASK. However, here we restrict the discussion to only single core comparison to evaluate the methods with respect to their basic FL operation costs.

Under these conditions,  elastic calculations took in average 12.13 seconds (measured over 100 calculations) while the AI-based approach only took 0.12 seconds (measured over 2000 calculations), revealing a speed-up of about 103 times, roughly corresponding to the reduction in the number of required FL operations.

The calculation for elastic-plastic constitutive material response required in DAMASK to reach the applied load in 100 increments took about 22 minutes, an average per time increment of 13.2 seconds, while the AI-based calculation took only 0.158 seconds (measured as an average over 2000 evaluations). This corresponds to a speed-up by a factor of about 8,300 times, corresponding to the same factor in reduced required total FL operations when using the AI-based solver for the application of the full load. Note that the simulation time for the non-linear solver in DAMASK will depend on the step size and numbers of increments applied. The current selection of 100 increments is not the optimum choice, however, based on the measurements, even calculation of a single increment for elasto-plastic material behaviour is about 84 times slower than the AI-based solver for a scenario where the total external load is applied that brings the material into the elastic-plastic regime. 

\section{Conclusions and Outlooks}
\label{sec_conclusion}

Advanced materials and the products consisting of them have become immensely complex in their internal structure and chemistry as well as in the mechanical loading conditions they get exposed to, ranging from cell phones to space ships. 
Particularly structural materials, i.e. materials  that   are primarily  meant to carry mechanical loads, nowadays contain multiple crystals, defects and phases with high mechanical contrast.
Mapping such complex microstructures and parts in the form of digital twins and subjecting them to mechanical calculations is a prerequisite to their microstructural design, improved processing, further development and safe application. Similar aspects apply to other mechanically heterogeneous media  as encountered in such diverse fields as soil-, structure-, building-, construction-, earthquake-, ice-, or colloidal mechanics.

Conducting such mechanical simulations for materials with  large internal phase contrast and complex constitutive  response (e.g. elasto-plastic behavior), together with non-linearities arising from large deformations, lead to  immense computational costs when using conventional finite-element or spectral solver methods. The high  computational costs are primarily due to the iterative nature of the solution algorithms used by these solvers. They limit the range, size and complexity of problems that are accessible to simulation-based investigations using current computers. 

Therefore, an alternative surrogate machine-learning-based solver for mechanically heterogeneous and non-linear fields (here stress distribution) is introduced in this work. The accuracy and computational costs of this U-net based solver are compared with a high-performance spectral solver, both for elastic and elasto-plastic constitutive  scenarios. The proposed deep neural network (DNN) can predict the local stress distribution with 3.8\% and 6.4\% mean absolute percentage error (MAPE) for heterogeneous elastic and elasto-plastic material response, respectively. The performance tests show a reduction in computation time of about 103 and 8300 times for elastic and elasto-plastic materials, respectively. Besides the acceptable accuracy and the substantially reduced  computational costs, the trained DNN also shows a great generalization capability by predicting stress distributions in geometries far different from those used in the training data-sets. 

Although the observed  MAPE shows a reasonably accurate stress prediction by the DNN method, a detailed error analysis and network architecture study reveals several pathways for its further improvement in the future. Incorporating filtering schemes to remove the high frequency noise and enforcing periodicity in kernel operations of the DNN represent specific items along these lines that are currently in  progress. In addition, predicting full stress and strain tensor components is straightforward based on the current framework and require no conceptual change.

The huge speed-up in calculation of the stress distribution in  such highly non-linear mechanical systems paves the way for many new applications, specifically addressing more complicated simulations, materials, loading scenarios  and optimization problems, not previously computationally accessible. Furthermore,   combining the proposed surrogate model with  conventional solution methods  can be an approach towards higher accuracy of  conventional solvers  at   considerably lower computational costs. This can be for instance achieved by employing the U-net based stress prediction as an initial stress guess for  conventional iterative solvers. This could greatly reduce the number of  iterations required, thus substantially saving computation time that is usually required for convergence.

\clearpage

\bibliographystyle{naturemag}
\bibliography{shorttitles, Ref}

\end{document}